# Hybrid MBE Route to Adsorption-Controlled Growth of BaTiO$_3$ Membranes with Robust Polarization Switching

Sooho Choo[1,*], Shivasheesh Varshney[1], Jay Shah[1], Anusha Kamath Manjeshwar[1], Dong Kyu Lee[1], K. Andre Mkhoyan[1], Richard D. James[2] and Bharat Jalan[1,*]

[1]Department of Chemical Engineering and Materials Science, University of Minnesota, Minneapolis, Minnesota, 55455, USA

[2]Department of Aerospace Engineering and Mechanics, University of Minnesota, Minneapolis, Minnesota, United States – 55455

* Corresponding authors: Sooho Choo, choo0021@umn.edu; Bharat Jalan, bjalan@umn.edu

**Abstract**

Freestanding ferroelectric membranes are promising for flexible electronics, nonvolatile memory, photonics, and spintronics, but their synthesis is challenged by the need for reproducibility with precise stoichiometric control. Here, we demonstrate the adsorption-controlled growth of single-crystalline, epitaxial BaTiO$_3$ films by hybrid molecular beam epitaxy (MBE) on a binary oxide sacrificial layer. Using a simple water-droplet lift-off method, we obtained submillimeter- to millimeter-sized membranes that retained crystallinity, as confirmed by high-resolution X-ray diffraction, and exhibited robust tetragonal symmetry by Raman spectroscopy. Impedance spectroscopy confirmed a high dielectric constant of ~1340, reflecting the robust dielectric response of the membranes. Ferroelectric functionality was revealed by piezoresponse force microscopy (PFM) and further verified by polarization–electric field (P-E) loop measurements with Positive-Up-Negative-Down (PUND). The P-E loops exhibited a remnant polarization of ~5 µC cm$^{-2}$ and a coercive field of ~63 kV cm$^{-1}$. These results were interpreted in relation to *c*- and *a*-domain configurations. These results establish hybrid MBE as a generalizable route for producing stoichiometry-controlled ferroelectric membranes, enabling their integration into next-generation flexible and multifunctional quantum oxide devices.

Keywords: adsorption-controlled growth, cation stoichiometry, freestanding oxide thin film, Raman spectroscopy, hybrid MBE, ferroelectricity

# 1. Introduction

Ferroelectric oxide thin films have been widely investigated for their nonvolatile polarization, which underpins low-power devices such as ferroelectric tunnel junctions and ferroelectric field-effect transistors[1-3]. Their functionality has recently been expanded by freestanding membrane forms, enabling integration with dissimilar material platforms[4], twist engineering[5, 6], and strain tuning[7-9], thereby offering new routes to tailor physical properties.

BaTiO$_3$ (BTO) is a prototypical lead-free high-$k$ dielectric and ferroelectric[10-12], exhibiting a tetragonal structure at room temperature (a = b = 3.992 Å, c = 4.036 Å) and a Curie temperature of 120 °C[13]. It also possesses one of the largest Pockels coefficients among perovskites[14, 15], making it a leading candidate for modulators, low-power optical switches, and high-speed photonic devices. Realizing these applications often requires integration of BTO with chemically and structurally dissimilar substrates, underscoring the need for freestanding membranes. Moreover, because the ferroelectric properties of BTO are highly sensitive to strain, stoichiometry, and oxygen vacancy concentration[16-19], synthesis strategies that yield freestanding membranes with high structural quality, precise stoichiometric control, and robust ferroelectricity are essential.

Conventional molecular beam epitaxy (MBE) approach has produced high-quality epitaxial BTO films[20-26], though the direct polarization-electric field (P-E) hysteresis has only demonstrated recently with remnant polarization of 15 μC cm$^{-2}$ in hybrid MBE-grown films[27]. Hybrid MBE utilizing volatile titanium tetraisopropoxide (TTIP), introduces an adsorption-controlled growth window where cation stoichiometry remains self-regulating[28-30]. Extending this approach to freestanding ferroelectric membranes introduces additional challenges: the heterostructure must incorporate a "MBE-friendly" sacrificial layer that is selectively etchable, thermally stable during growth, and chemically compatible to suppress interfacial reactions or diffusion[31, 32].

In this study, we demonstrate hybrid MBE growth of stoichiometry-controlled BTO membranes using optimized TTIP/Ba beam equivalent pressure (BEP) ratios. Adsorption-controlled growth on a SrO sacrificial layer[33] was verified by analyzing the out-of-plane lattice parameters. A water-droplet assisted lift-off enabled transfer of single-crystalline BTO membranes with preserved tetragonal symmetry, as confirmed by Raman spectroscopy. The ferroelectric functionality of these membranes was first probed by piezoresponse force microscopy (PFM), revealing local switchable polarization domains, and further established by P-E loop measurements with Positive-Up-Negative-Down (PUND). The P-E loops exhibited a remnant polarization of ~5 μC cm$^{-2}$ and a coercive field of ~63 kV cm$^{-1}$, revealing the role of $c$- vs. $a$-domain contributions.

## 2. Results and discussion

**Structural properties and growth optimization of BaTiO$_3$ films on SrO/LSAT (001).**

A 60 nm BTO film was synthesized on a 10-15 nm SrO sacrificial layer grown on an LSAT (001) substrate (Figure 1a) at 950 °C using hybrid MBE. Sr and Ba were evaporated from solid-source effusion cells, while Ti was supplied from TTIP, introduced through a vapor inlet system. The fluxes of Sr, Ba, and TTIP were calibrated as BEPs using an ion gauge positioned in the molecular beam path. Additional growth details are provided in the Experimental Section. Because the face-diagonal ($a/\sqrt{2}$ = 3.64 Å) of SrO layer[34] is better lattice-matched to that of the edge of BTO, a 45° in-plane rotation between the (001) SrO sacrificial layer and (001) BTO is expected to facilitate epitaxial stabilization. Consistently, Figure 1b shows an X-ray diffraction (XRD) in-plane ϕ-scan of a 60 nm BTO/15 nm SrO/LSAT (001) around the (200) reflection revealing an epitaxial relationship BTO [100]∥SrO[110]∥LSAT[100] and BTO (001)∥SrO (001)∥LSAT (001). Figure 1c further shows the reciprocal space map (RSM) for BTO (103) and LSAT (103), revealing that the BTO film is nearly fully relaxed on the SrO sacrificial layer.

To optimize growth conditions for stoichiometric BTO films on a SrO sacrificial layer, we deposited 60 nm BTO films on 10 nm SrO/LSAT (001) substrates using hybrid MBE, systematically varying the TTIP/Ba BEP ratio from 73 to 183 by adjusting the Ba flux at a fixed TTIP flux. Figure 1d shows high-resolution (HR)XRD $2\theta$-$\omega$ scans of the resulting films. For the BTO film grown at a BEP ratio of 73, the SrO layer exhibited a reduced out-of-plane lattice parameter of 5.106 Å compared to the bulk value ($a_{sro}$, bulk = 5.139 Å), along with markedly suppressed intensity. This indicates instability of the SrO layer under Ba-rich conditions. Moreover, the corresponding BTO film was polycrystalline, showing both (002) and (101) peaks in HRXRD, consistent with the spotty ring patterns observed in the reflection high-energy electron diffraction (RHEED) image after growth (Figure S1, Supplementary Information). In contrast, films grown with TTIP/Ba BEP ratios between 88 and 183 displayed epitaxial SrO with a lattice parameter of 5.148 ± 0.002 Å and single BTO (*00l*) peaks, confirming phase-pure, single-crystalline epitaxial films.

As an indirect measure of film stoichiometry, the out-of-plane lattice parameter was extracted from the BTO (002) peak (Figure 1e). At a BEP ratio of 73, the lattice parameter was reduced relative to the bulk c-axis value (3.974 Å vs. 4.036 Å), indicating nonstoichiometric BTO films, likely owing to an intermixing with the SrO layer, or from complex defect formation associated with the Ba-rich growth conditions. At a BEP ratio of 88, the c-axis expanded to 4.022 Å, still below the bulk value, marking the boundary of stoichiometric growth close to the Ba-rich regime. For TTIP/Ba BEP ratios between 96 and 155, the BTO films exhibited nearly constant c-axis values of 4.040 ± 0.003 Å, in close agreement with the bulk (4.036 Å). This plateau indicates an adsorption-controlled growth regime, where excess volatile Ti-containing species desorb to maintain cation stoichiometry[35]. However, at the highest BEP ratio (183), the c-axis expanded to 4.086 Å, consistent with Ba deficiency and associated cation nonstoichiometry[36]. Films grown outside the stoichiometric window also exhibited degraded structural quality, evidenced by increased root mean square (RMS) surface roughness measured using atomic force microscopy (AFM) and broadening of the full-width-half-maxima (FWHM) of BTO (002) rocking curve (Figures S1 and S2, Supporting Information).

**Transfer process and structural properties of BaTiO$_3$ membranes**

BTO films were exfoliated by dissolving the SrO sacrificial layer, and the resulting freestanding BTO membranes were transferred onto Au-coated Si substrates, as illustrated in Figure 2a. During transfer, the as-grown films were flipped such that the BTO surface faced downward onto the Au layer. The SrO layer was then dissolved in a water droplet, and upon evaporation the LSAT substrate detached, leaving the BTO membranes on the Au-coated Si. Details of the transfer process are provided in the Experimental Section. Exfoliation was unsuccessful under Ba-rich conditions (TTIP/Ba BEP ratio = 73) due to SrO degradation, consistent with the diminished SrO (002) intensity observed in XRD (Figure 1d). Under Ba-deficient conditions (TTIP/Ba BEP ratio = 183), only small fractions of flat BTO membranes were obtained (Figure S3, Supporting Information), suggesting that cation stoichiometry may also be influencing the mechanical robustness of the membranes.

The structure of the exfoliated membranes was examined by HRXRD. A representative HRXRD scan of a stoichiometric BTO membrane (TTIP/Ba BEP ratio = 113) is shown in Figure 2b, exhibiting phase-pure, single-crystalline diffraction peaks with bulk-like lattice parameters. Results for other membranes are provided in Figure S4 (See Supporting Information). The out-of-plane lattice parameters extracted from the BTO (002) peaks of transferred membranes (solid blue symbols) followed the same stoichiometry-dependent trend as the as-grown films (solid gray symbols) (Figure 2c). A slight reduction in lattice parameter was consistently observed across all membranes, which may arise from the absence of substrate clamping or an increased density of *a*-domains (discussed later). To probe atomic structure, a 20 nm stoichiometric membrane (TTIP/Ba BEP ratio = 109) was transferred onto a Cu TEM grid, where optical imaging (Figure 2d) showed coverage over hundreds of micrometers. Plan-view high-angle annular dark field scanning transmission electron microscopy (HAADF-STEM) (Figure 2e) revealed well-ordered Ba and Ti atomic columns, confirming the single-crystalline nature of the membrane.

**Structural properties of BTO membranes characterized by Raman spectroscopy**

Because the tetragonal distortion of BTO is closely linked to its ferroelectric properties, we performed Raman spectroscopy on BTO membranes grown with TTIP/Ba BEP ratios ranging from 88 to 183 to probe their tetragonal symmetry and structural order (Figure 3). The vertical dashed lines in Figure 3 mark three representative Raman modes of tetragonal BTO: 305 cm$^{-1}$ [$B_1$ and E(TO+LO)], 525 cm$^{-1}$ [$A_1$(TO) and E(TO)], and 715 cm$^{-1}$ [$A_1$(LO) and E(LO)][37]. The $B_1$ mode arises from oxygen vibrations in $BO_6$ octahedra, E modes correspond to atomic vibrations in the a-b plane, and $A_1$ modes are associated with atomic motions along the c-axis[38].

For Ba-deficient membrane (TTIP/Ba BEP = 183), two broad features near 250 cm$^{-1}$ and 525 cm$^{-1}$ were observed, likely second-order Raman modes, indicating substantial structural disorder in the tetragonal lattice[39]. In contrast, stoichiometric membranes exhibited clear tetragonal

Raman signatures. Within the stoichiometric growth window (88 ≤ TTIP/Ba BEP ≤ 183), although all Raman modes were observed, the lower TTIP/Ba BEP ratios produced sharper Raman peaks, reflecting stronger tetragonal symmetry[40]. These variations across stoichiometric membranes likely reflect differences in microstructure of membranes, consistent with rocking curve results from the as-grown films (Figure S2, Supporting Information). Overall, Raman spectroscopy confirms robust tetragonal symmetry in all membranes BTO membranes within the MBE-grown window.

**Ferroelectric properties of BaTiO$_3$ membranes**

Dielectric and ferroelectric properties of stoichiometric BTO membranes were investigated using PFM, impedance spectroscopy and P-E loop measurements. Figure 4a shows the PFM setup and an optical micrograph of a 20-nm-thick BTO membrane grown within the growth window (TTIP/Ba BEP ratio = 100). After applying -10 V to an outer box (6 µm × 6 µm) and +10 V to an inner box (3 µm × 3 µm) using a conductive tip, stable out-of-plane domains were written, as evidenced by the phase and amplitude images in Figures 4b and 4c, respectively. The persistence of the written domains demonstrates robust polarization retention. Furthermore, local PFM spectroscopy revealed a 180° phase contrast and a butterfly-shaped amplitude loop, confirming reversible polarization switching (Figure S5, Supporting Information).

Finally, to investigate dielectric and ferroelectric behavior, we fabricated a metal-insulator-metal (MIM) capacitor with Pt electrodes (Figure 4d). A 220 nm Pt film was deposited on BTO films prior to exfoliation and transfer (Experimental Section and Figure S6, Supporting Information). This Pt supporting layer enabled a crack-free, millimeter-sized BTO membrane but also introduced macroscopic wrinkles (Figure S6, Supporting Information) and prevented post-growth oxygen annealing - necessary for mitigating oxygen vacancies - due to the use of double-sided Kapton tape in the transfer process (Figure 4d schematic). Figure 4e shows the reactance magnitude (Z") (solid blue symbol) and the phase angle (solid red symbol) as a function of frequency. The phase angle remained close to -90° between 100 Hz and 2 kHz, characteristic of ideal capacitive behavior, but decreased at higher frequencies due to leakage, likely associated with oxygen vacancies. The capacitance was extracted from a linear fit of the reactance magnitude (|Z"|) in the range of 100 Hz and 2 kHz where the phase angle approached -90°, using the relation:

$$\log(|Z''|) = -\log(2\pi f) - \log(C).$$

For the MIM structure, the static dielectric constant ($\varepsilon_r$) was obtained from

$$\varepsilon_r = \frac{C\, t_{BTO}}{\varepsilon_0\, A}$$

where $t_{BTO}$ is the BTO membrane thickness, $\varepsilon_0$ is the permittivity of free space, and $A$ is the top electrode area. The fitting yielded a dielectric constant of ~1340 (Figure 4e), which lies between the reported values for bulk BTO along the c-axis (~200) and a-axis (~4000)[41]. Although the exact origin of this enhanced dielectric response is not fully clear, it can be qualitatively attributed to the coexistence of a- and c-domains, likely promoted by wrinkle formation. This interpretation is consistent with the measured out-of-plane lattice parameter of 4.018 Å, which falls between the lattice parameters of the a- and c-domains of BTO (Figure S6).

After confirming the capacitive regime, P-E loop measurements were performed at 300 K using PUND at 2 kHz (Figure 4f; see Experimental Section and Figure S7, Supporting Information). This approach effectively eliminates contributions from non-ferroelectric and leakage currents. The BTO membranes exhibited the $P_r$ of ~5 µC cm$^{-2}$ and a coercive field ($E_c$) of ~63 kV/cm. The $P_r$ value is smaller than that of bulk BTO likely due to the coexistence of a- and c-domains, consistent with the high dielectric constant[42, 43]. Together, these results hint at the potential for domain engineering in BTO membranes through control of strain and wrinkle formation, while also demonstrating that hybrid MBE enables adsorption-controlled growth of freestanding BTO membranes that preserve both structural integrity and robust ferroelectric functionality.

## 3. Conclusion

In this study, we report a hybrid MBE route to synthesize and transfer stoichiometric ferroelectric BTO membranes by leveraging an adsorption-controlled growth window with volatile TTIP precursors. Water-assisted exfoliation of SrO sacrificial layers produced submillimeter-to-millimeter membranes that preserve crystallinity, tetragonal symmetry, and robust ferroelectricity. This scalable approach not only overcomes long-standing challenges in creating freestanding perovskite with self-regulating stoichiometry control but also establishes BTO membranes as a powerful platform for strain engineering, twisted-membrane physics, and 2D/3D heterointegration toward energy-efficient and neuromorphic technologies.

## 4. Experimental section

### BaTiO$_3$ membrane growth and characterizations

BTO films and SrO sacrificial layers were grown on 5 mm × 5 mm LSAT (001) substrates using hybrid MBE[29]. Substrates were cleaned sequentially in acetone, methanol, and isopropyl alcohol (IPA) for 5 min each, then loaded into a fast-entry load-lock (FEL) chamber and baked at 150 °C for 3 h before transfer to the growth chamber. Film growth was performed at 950 °C (substrate temperature measured by thermocouple) in an MBE system equipped with effusion cells for solid sources (Sr, Ba), a metal–organic inlet system for TTIP, and an inductively coupled RF plasma source for oxygen. Prior to deposition, oxygen plasma cleaning was carried out at 250 W for 15 min.

For the SrO sacrificial layer, growth was conducted with a Sr BEP of ~$1 \times 10^{-7}$ mbar and an oxygen partial pressure of $5 \times 10^{-6}$ mbar. BTO films were subsequently deposited with Ba BEPs ranging from $6.8 \times 10^{-8}$ to $1.43 \times 10^{-7}$ mbar and a fixed TTIP Baratron pressure of 140 mTorr (corresponding to a TTIP flux of $1 \times 10^{-5}$ mbar). Surface crystallinity was monitored in situ using RHEED (Staib Instruments). Structural characterization of the films was carried out by HRXRD, RSM, and in-plane ϕ-scans using a Smart Lab XE diffractometer (Rigaku). Surface topography and roughness were examined by AFM (Nanoscope V Multimode 8, Bruker).

Transferred BTO membranes on Au-coated Si substrates were inspected by optical microscopy (Optiphot, Nikon) with a digital camera (SL1, Canon). Raman spectra were collected using a confocal Raman microscope (Alpha 300R, WITec). Ferroelectric polarization of BTO membranes was probed by PFM (Dimension ICON, Bruker) using conductive tips (SCM-PTSI, Bruker).

STEM images were obtained using an aberration-corrected FEI Titan G2 60-300 (S)TEM equipped with a CEOS DCOR probe corrector. The microscope was operated at 200kV with probe convergence angle of 25.5 mrad and a beam current of 50pA. The camera length was set at 130mm, with the HAADF detector inner and outer collection angles of 55 and 200 mrad, respectively.

**Transfer process by dissolving sacrificial layers with a water droplet**

The as-grown BTO film was placed face down on the host substrate. A droplet of deionized (DI) water was applied with a pipette to fully cover the BTO/SrO/LSAT stack and left for 10 min to dissolve the SrO sacrificial layer. Once the layer was completely dissolved, the sample was heated on a hot plate at 60 °C to evaporate the DI water. The LSAT substrate was then carefully lifted off with tweezers, leaving the BTO membrane transferred onto the host substrate.

**Device fabrication process**

A 220 nm Pt layer was first deposited by DC sputtering (ATC 2000, AJA International) on the 240 nm as-grown BTO film to serve as a mechanical support. The Pt/BTO/SrO/LSAT stack was then placed face down on a Si substrate coated with $SiO_2$ and double-sided Kapton tape. The SrO sacrificial layer was dissolved in DI water over 12 h, transferring a 5 mm × 5 mm BTO membrane onto the host substrate with a continuous Pt bottom electrode.

Subsequently, circular Pt top electrodes (50 μm diameter) were fabricated by photolithography. A photoresist (NR71 3000P) was spin-coated at 3000 rpm for 45 s, soft-baked at 150 °C for 60 s, exposed to UV light for 20 s in hard-contact mode, and post-baked at 100 °C for 60 s. The unexposed regions were removed with RD6 developer, leaving hardened photoresist on the patterned areas. Finally, 110 nm Pt was deposited by DC sputtering to form the top electrodes, followed by liftoff of the photoresist.

**P-E loop measurement with PUND**

PUND data of the $BaTiO_3$ membrane was obtained using a waveform generator module on a semiconductor analyzer (B1500A, Keysight) at a frequency of 2 kHz at room temperature. For PUND measurements, the pulse train has the rise (fall), wait, and delay time is 125 µs, 1 µs, and 100 ns, respectively. The voltage was applied via top electrodes, and bottom electrodes were used as a ground.

**Acknowledgements**

Film growth (S.C. and S.V.) was supported primarily by the U.S. Department of Energy (Award No. DE-SC0020211) and in part by the Center for Programmable Energy Catalysis, an Energy Frontier Research Center funded by the U.S. Department of Energy, Office of Science, Basic Energy Sciences at the University of Minnesota, under Award No. DE- SC0023464. This work also benefitted from the Air Force Office of Scientific Research Multi University Research Initiative (AFOSR MURI, Award No. FA9550-25-1-0262). Film growth was performed using instrumentation funded by AFOSR DURIP awards FA9550-18-1-0294 and FA9550-23-1-0085. J.S and K.A.M. would also like to acknowledge support from the National Science Foundation through the Materials Research Science and Engineering Center (MRSEC, Award No. DMR-2011401). Parts of this work were carried out in the College of Science and Engineering Characterization Facility at the University of Minnesota which receives partial support MRSEC and the National Nanotechnology Coordinated Infrastructure (NNCI, Award No. ECCS-2025124) programs. S.V. also acknowledges the University of Minnesota Doctoral Dissertation Fellowship.

**Author contributions**

S.C., R.D.J., and B.J. conceived the idea and designed the experiments. S.C. grew the thin film heterostructures. S.C. exfoliated and transfer membranes with assistance from S.V. and D.K.L. S.C. characterized the structural properties using X-ray diffraction and reciprocal space mapping. S.C. fabricated the defined capacitors and performed impedance spectroscopy. S.C. and A.K.M.

performed frequency-dependent four-quadrant current vs. voltage measurements and performed the PUND and static *I-V* measurements and analyzed the data. J.S. performed STEM imaging under the supervision of K.A.M. S.C. and B.J. wrote the manuscript. All authors contributed to the discussion of the results and reviewed the manuscript.

**Conflict of interest**

The authors declare that they have no conflict of interest.

**Data availability**

All data needed to evaluate the conclusions of the paper are present in the paper and/or the Supplementary Information.

**Supplementary information**

Details on surface morphology, rocking curves of as-grown films, optical microscopy and structural characterization of membranes as a function of stoichiometry, local PFM loops, device fabrication steps, and optical/AFM images of the final device structure are provided in the Supplementary Information.

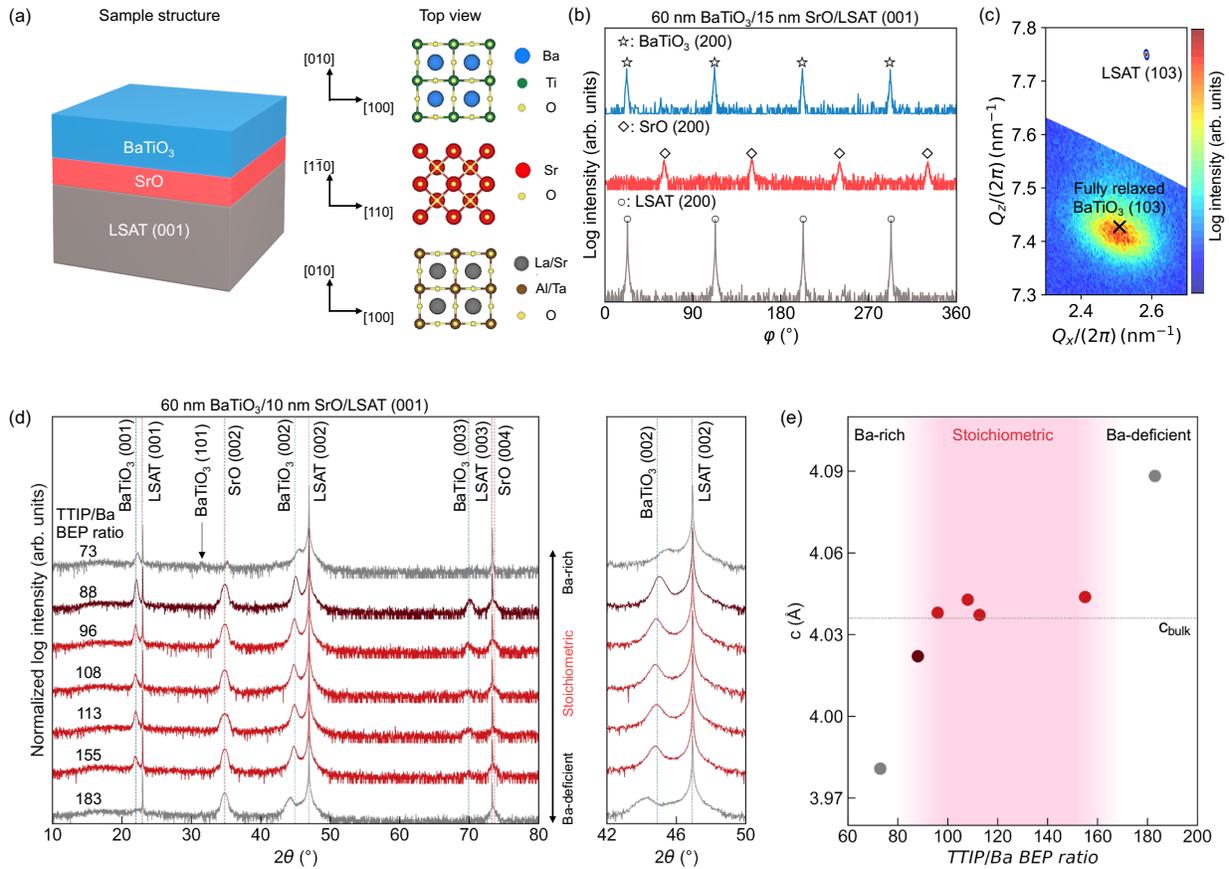

**Figure 1. Structural properties of as-grown BaTiO$_3$ films with a SrO sacrificial layer on an LSAT (001) substrate**. (a) Schematic illustration of sample structure, including the functional layer (BaTiO$_3$), the sacrificial layer (SrO), and the substrate (LSAT (001)), along with top view of atomic crystal structure of BaTiO$_3$ (perovskite), SrO (rock salt), and LSAT (perovskite). (b) In-plane $\phi$-scan of a 60 nm BaTiO$_3$/15 nm SrO/LSAT (001) structure for the (200) plane, showing that the BaTiO$_3$, SrO, and LSAT have four-fold symmetry, and the SrO (200) was rotated by nearly 45º with respect to the BaTiO$_3$ (200) and LSAT (200). (c) RSM of the BaTiO$_3$ layer for the BaTiO$_3$ and LSAT (103) reflections, indicating a mostly relaxed BaTiO$_3$ film. (d) HRXRD $2\theta$-$\omega$ coupled scans for a 60 nm BaTiO$_3$/10 nm SrO/LSAT (001) structure grown at various TTIP/Ba BEP ratios ranging from 73 to 183 (varying Ba flux). The wide scans are shown on the left, and the finer range of $2\theta$ for BaTiO$_3$ (002) reflection are displayed on the right. Blue, red, and gray dashed vertical lines refer to XRD reflections for BaTiO$_3$ (00$l$), SrO (00$l$), and LSAT (00$l$), respectively. (e) Out-of-plane lattice parameter as a function of the TTIP/Ba BEP ratio for as-grown BaTiO$_3$ films, calculated from the results of BaTiO$_3$ (002) peak in HRXRD $2\theta$-$\omega$ coupled scans.

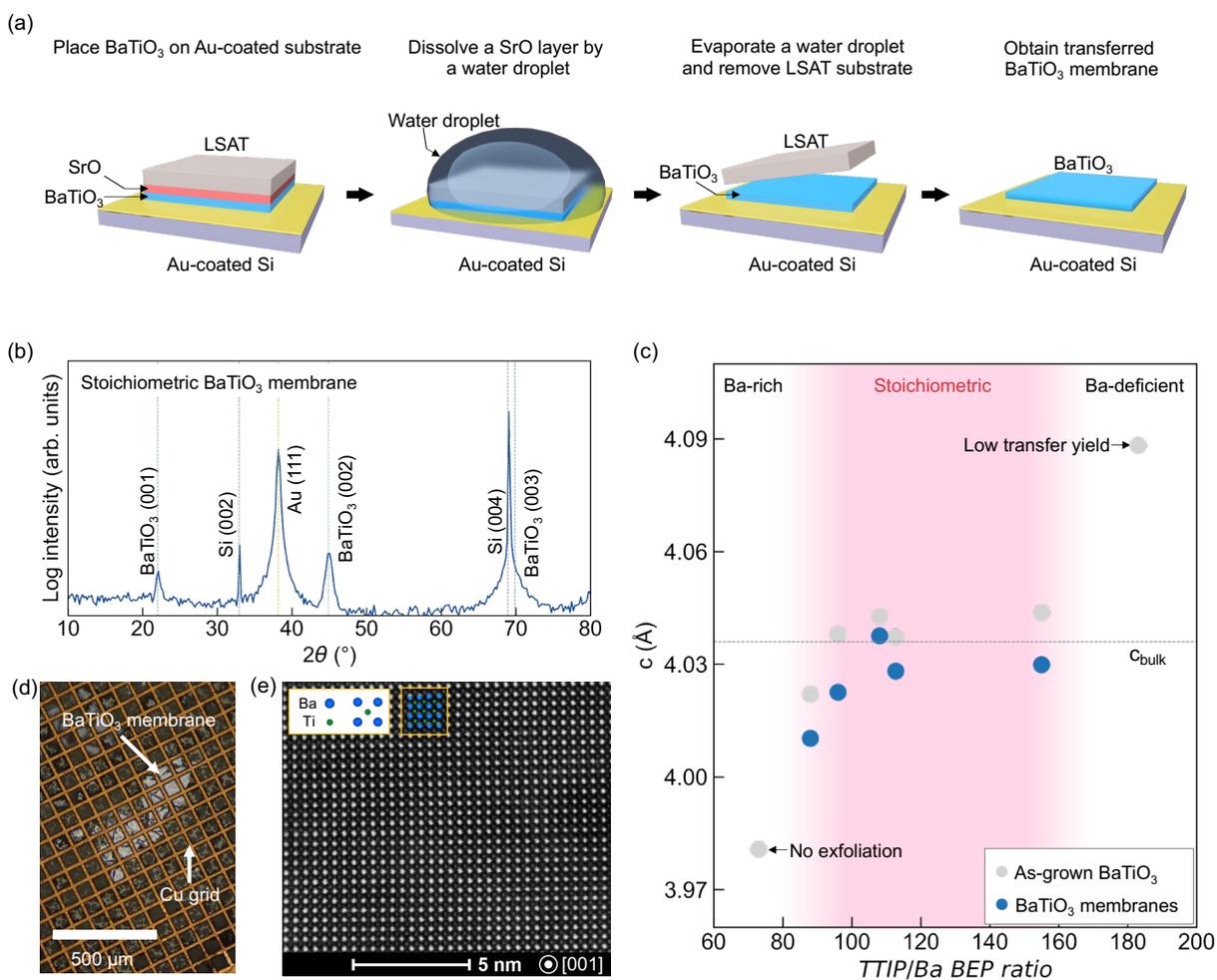

**Figure 2. Transfer process and structural characterization of BaTiO₃ membranes**. (a) Schematic illustration of the transfer process: 1) Place the BaTiO$_3$ film on an Au-coated substrate, 2) Dissolve the SrO sacrificial layer using a water droplet, 3) Evaporate the water droplet and remove the LSAT substrate, and 4) Obtain the freestanding BaTiO$_3$ membrane. (b) Representative HRXRD 2$\theta$-$\omega$ coupled scan of a stoichiometric BaTiO$_3$ membrane grown at a TTIP/Ba BEP ratio of 113 on an Au-coated Si substrate. (c) Out-of-plane lattice parameters as a function of the TTIP/Ba BEP ratio for BaTiO$_3$ membranes, calculated from HRXRD 2$\theta$-$\omega$ coupled scans. Blue circles for BaTiO$_3$ membranes and gray circles for as-grown BaTiO$_3$ films. (d) Optical micrograph of 20 nm BaTiO$_3$ membrane on a Cu grid. (e) HAADF-STEM image of the 20 nm BaTiO$_3$ membrane. Ba and Ti atoms represented as blue and green circles in the yellow box, respectively.

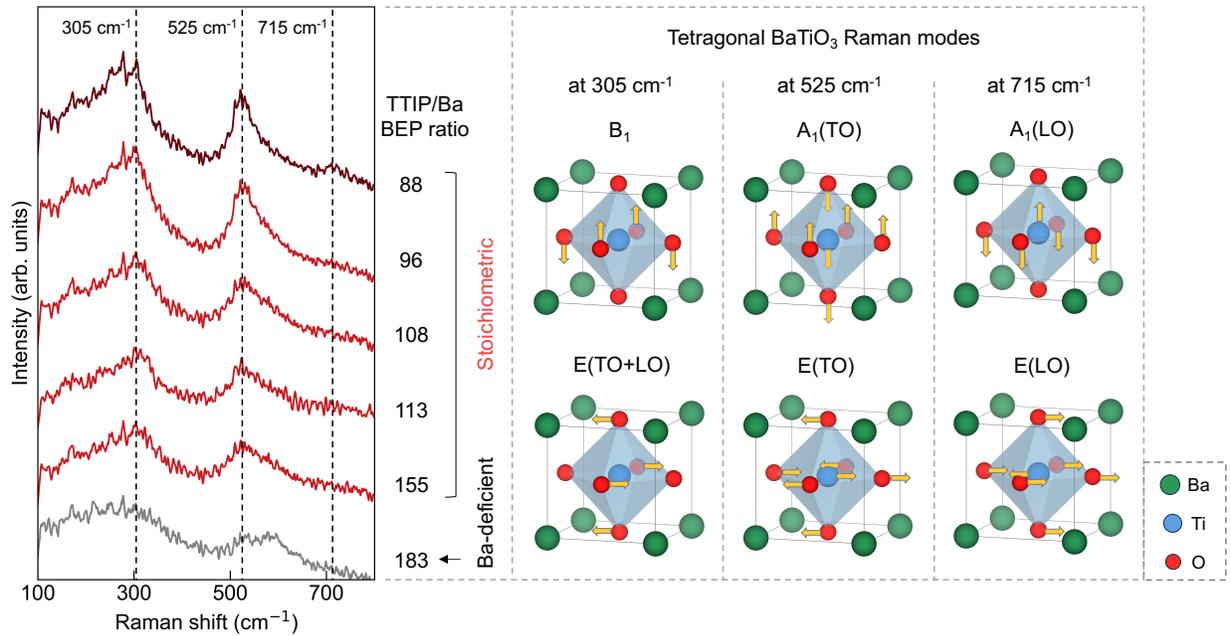

**Figure 3. Structural properties of BaTiO$_3$ membranes characterized by Raman spectroscopy**. Raman spectra of BaTiO$_3$ membranes grown under TTIP/Ba BEP ratios from 88 to 183 with schematic illustrations of Raman modes of tetragonal BaTiO$_3$. Dashed lines indicate the Raman modes at 305 cm$^{-1}$, 525 cm$^{-1}$ and 715 cm$^{-1}$. B$_1$ and E(TO+LO) modes are shown at 305 cm$^{-1}$. A$_1$(TO) and E(TO) modes present at 525 cm$^{-1}$. A$_1$(LO) and E(LO) modes are observed at 715 cm$^{-1}$. Green, blue, and red circles indicate Ba, Ti, and O atoms, respectively, and yellow arrows refer to direction of phonon vibrations in the schematic illustrations.

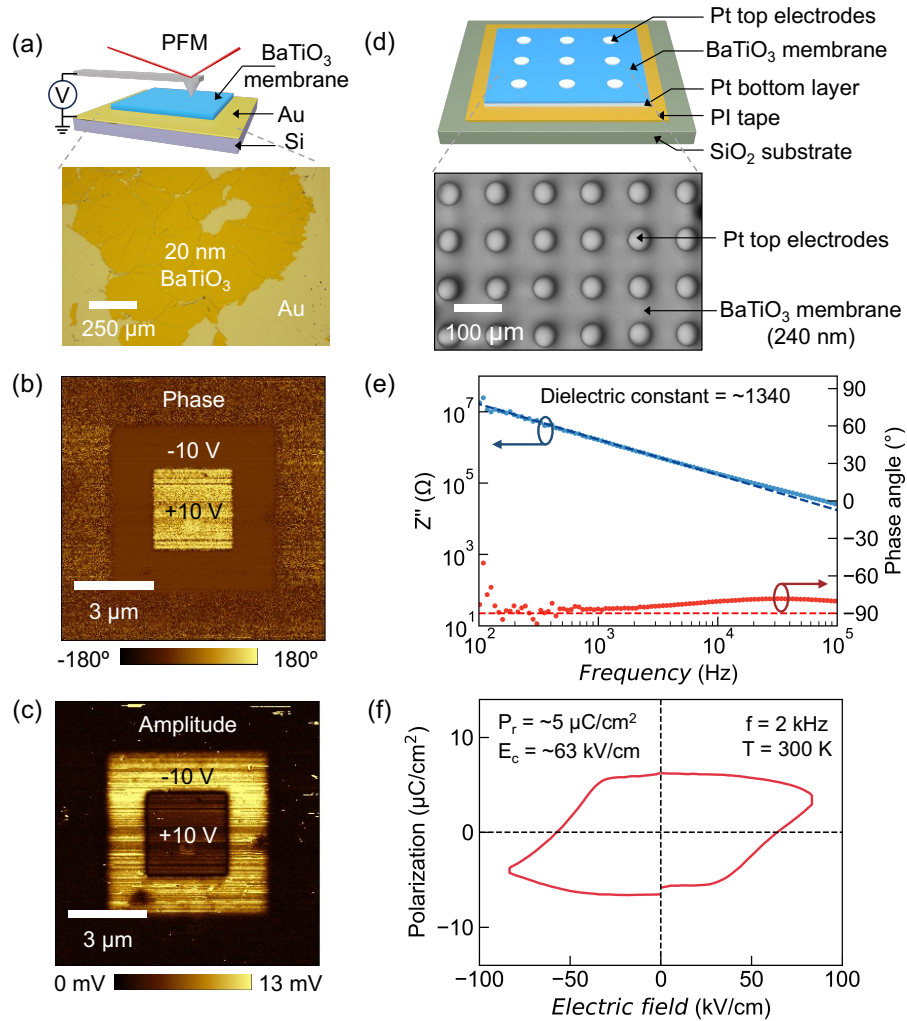

**Figure 4. Ferroelectric properties of stoichiometric BaTiO₃ membranes**. (a) Schematic illustration of PFM set up with an optical micrograph of the 20 nm BaTiO$_3$ membrane. PFM out-of-plane (b) phase image and (c) amplitude image of the BaTiO$_3$ membrane written with ±10 V DC bias. (d) Schematic illustration of the MIM structure (Pt/BaTiO$_3$/Pt) attached on a SiO$_2$ substrate using double-sided Kapton tape with an optical micrograph of the BaTiO$_3$ device with Pt top electrodes. (e) Imaginary part of impedance and phase angle of the BaTiO$_3$ membrane as a function of frequency at room temperature. The blue dashed line is the linear fit to the impedance imaginary part in the frequency range between 100 Hz and 2 kHz, where the phase angle close to -90º and red dashed line indicates the phase angle of -90º. (f) P-E hysteresis loops of the BaTiO$_3$ membrane measured using the PUND method (frequency = 2 kHz, room temperature).